\documentclass[a4paper]{article}
\usepackage{cite}
\usepackage{INTERSPEECH2019}
\usepackage{float}
\usepackage{subfigure}
\usepackage{graphicx}
\usepackage{multirow}
\usepackage{booktabs}
\usepackage{tabularx}
\usepackage{hyperref}
\makeatletter
\newcommand{\printfnsymbol}[1]{%
  \textsuperscript{\@fnsymbol{#1}}%
}
\makeatother
\title{DCCRN: Deep Complex Convolution Recurrent Network for Phase-Aware Speech Enhancement}
\name{Yanxin Hu$^{1,*}$\thanks{*: Equal contribution. The first author performed part of this work as an intern at Sogou. Lei Xie is corresponding author.}, Yun Liu$^{2,*}$, Shubo Lv$^1$, Mengtao Xing$^1$, Shimin Zhang$^1$,\\  Yihui Fu$^1$, Jian Wu$^1$, Bihong Zhang$^2$, Lei Xie$^1$}
\address{
  $^1$Audio, Speech and Language Processing Group (ASLP@NPU), School of Computer Science, Northwestern Polytechnical University, Xi\textquotesingle an, China\\
  $^2$AI Interaction Division, Sogou Inc., Beijing, China}
\email{\{yxhu, shblv, mtxing, shmzhang, yhfu\}@npu-aslp.org, \{jianwu,lxie\}@nwpu-aslp.org, liuyun.nogizaka@qq.com, zhangbihong@sogou-inc.com}

\begin{document}
\renewcommand{\thefootnote}{}
\newcommand{\tabincell}[2]{\begin{tabular}{@{}#1@{}}#2\end{tabular}}
\setlength{\abovecaptionskip}{0pt} 
\setlength{\belowcaptionskip}{0pt}
\setlength{\abovedisplayskip}{1pt}
\setlength{\belowdisplayskip}{1pt}

\maketitle
\begin{abstract}
Speech enhancement has benefited from the success of deep learning in terms of intelligibility and perceptual quality. Conventional time-frequency (TF) domain methods focus on predicting TF-masks or speech spectrum, via a naive convolution neural network (CNN) or recurrent neural network (RNN).  Some recent studies use complex-valued spectrogram as a training target but train in a real-valued network, predicting the magnitude and phase component or real and imaginary part, respectively. Particularly, convolution recurrent network (CRN) integrates a convolutional encoder-decoder (CED) structure and long short-term memory (LSTM), which has been proven to be helpful for complex targets. In order to train the complex target more effectively, in this paper, we design a new network structure simulating the complex-valued operation, called Deep Complex Convolution Recurrent Network (DCCRN), where both CNN and RNN structures can handle complex-valued operation. The proposed DCCRN models are very competitive over other previous networks, either on objective or subjective metric. With only 3.7M parameters, our DCCRN models submitted to the Interspeech 2020 Deep Noise Suppression (DNS) challenge ranked first for the real-time-track and second for the non-real-time track in 
terms of Mean Opinion Score (MOS). 

\end{abstract}
\hyphenpenalty=5000
\tolerance=1000
\section{Introduction}
Noise interference may severely decrease perceptual quality and intelligibility in speech communication. Likewise, the related tasks, such as automatic speech recognition (ASR), also can be heavily affected by noise interference. \textit{Speech enhancement} is thus a highly desired task of taking noisy speech as input and producing an enhanced speech output for better speech quality, intelligibility, and sometimes better criterion in downstream tasks (e.g., lower error rate in ASR). Recently, deep learning (DL) methods have achieved promising results in speech enhancement, especially in dealing with non-stationary noises in challenging conditions. DL can benefit both single-channel (monaural) and multi-channel speech enhancement depending on specific applications. In this paper, we focus on DL-based single-channel speech enhancement for better perceptual quality and intelligibility, particularly targeting to real-time processing with low model complexity. The Interspeech 2020 deep noise suppression (DNS) challenge has provided a common testbed for such purpose~\cite{reddy2020interspeech}.


\subsection{Related work}
Formulated as a supervised learning problem, noisy speech can be enhanced by neural networks either in time-frequency (TF) domain or directly in time-domain. The time-domain approaches can further fall into two categories --- direct regression~\cite{fu2018end,stoller2018wave} and adaptive front-end approaches~\cite{luo2019conv,luo2019dual,zhang2020furcanext}. The former directly learns a regression function from the waveform of a speech-noise mixture to the target speech without an explicit signal front-end, typically by involving some form of 1-D convolutional neural network (Conv1d). Taking time-domain signal in and out, the latter adaptive front-end approaches usually adopt a convolution encoder-decoder (CED) or 
a u-net framework, which resembles the short-time Fourier transform (STFT) and its inversion (iSTFT). The enhancement network is then inserted between the encoder and the decoder, typically by using networks with the capacity of temporal modeling, such as temporal convolutional network (TCN)~\cite{luo2019conv,bai2018empirical} and long short-term memory (LSTM)~\cite{weninger_erdogan_watanabe_vincent_roux_hershey_schuller_2015}. 

As another main-stream, the TF-domain approaches~\cite{srinivasan2006binary,narayanan2013ideal,zhao2016dnn,xu2013experimental,yin2019phasen} work on the spectrogram with the belief that fine-detailed structures of speech and noise can be more separable with TF representations after STFT. Convolution recurrent network (CRN)~\cite{tan2018convolutional} is a recent approach that also employs a CED structure similar to the one in the time-domain approaches but extracts high-level features for better separation by 2-D CNN (Conv2d) from noisy speech spectrogram. Specifically, CED can take complex-valued or real-valued spectrogram as input. A complex-valued spectrogram can be decomposed into magnitude and phase in polar coordinate or real and imaginary part in the Cartesian coordinate. For a long time, it has been believed that phase is intractable to estimate. Hence, early studies only focus on magnitude related training target while ignoring phase~\cite{huang2014deep,xu2014regression,takahashi2018mmdenselstm}, resynthesizing the estimated speech by simply applying estimated magnitude with the noisy speech phase. This thus limits the upper bound of performance, while the phase of estimated speech will deviate significantly with serious interferences. Although many recent approaches have been proposed for phase reconstruction to address this issue~\cite{wang2015deep,liu2019supervised}, the neural network remains real-valued.

Typically, training targets defined in the TF domain mainly fall into two groups, i.e., masking-based targets, which describe the time-frequency relationships between clean speech and background noise, and mapping-based targets which correspond to the spectral representations of clean speech. In the masking family, ideal binary mask (IBM)~\cite{wang2005ideal}, ideal ratio mask (IRM)~\cite{narayanan2013ideal} and spectral magnitude mask (SMM)~\cite{wang2014training} only use the magnitude between clean speech and mixture speech, ignoring the phase information. On the contrast, phase-sensitive mask (PSM)~\cite{erdogan2015phase} was the first one that utilizes phase information showing the feasibility of phase estimation. Subsequently,  complex ratio mask (CRM)~\cite{williamson2015complex} was proposed, which can reconstruct speech perfectly by enhancing both real and imaginary components of the division of clean speech and mixture speech spectrogram simultaneously. Later, Tan \textit{et al.}~\cite{tan2019complex} proposed a CRN with one encoder and two decoders for complex spectral mapping (CSM) to estimate the real and imaginary spectrogram of mixture speech simultaneously. It is worth noting that CRM and CSM possess the full information of a speech signal so that they can achieve the best oracle speech enhancement performance in theory.  

The above approaches have been learned under a real-valued network, although the phase information has been taken into consideration. Recently, deep complex u-net~\cite{choi2019phase} has combined the advantages of both a deep complex network~\cite{trabelsi2017deep} and a u-net~\cite{ronneberger2015u} to deal with complex-valued spectrogram. Particularly, DCUNET is trained to estimate CRM and optimizes the scale-invariant source-to-noise ratio (SI-SNR) loss~\cite{luo2019conv} after transforming the output TF-domain spectrogram to a time-domain waveform by iSTFT. While achieving state-of-the-art performance with temporal modeling ability, many layers of convolution are adopted to extract important context information, leading to large model size and complexity, which limits its practical use in efficiency-sensitive applications.

 
 \subsection{Contributions}
 
In this paper, we build upon previous network architectures to design a new complex-valued speech enhancement network, called \textit{deep complex convolution recurrent network} (DCCRN), optimizing an SI-SNR loss. The network effectively combines both the advantages of DCUNET and CRN, using LSTM to model temporal context with significantly reduced trainable parameters and computational cost. Under the proposed DCCRN framework, we also compare various training targets and the best performance can be obtained by the complex network with the complex target. In our experiments, we find that the proposed DCCRN outperforms CRN~\cite{tan2019complex} by a large margin.  With only 1/6 computation complexity, DCCRN achieves competitive performance with DCUNET~\cite{choi2019phase} under the similar configuration of model parameters. While targeting to real-time speech enhancement, with only 3.7M parameters, our model achieves the best MOS in real-time track and the second-best in non-real-time track according to the P.808 subjective evaluation in the DNS challenge.


\section{The DCCRN Model}
\subsection{Convolution recurrent network architecture}
The convolution recurrent network (CRN), originally described in~\cite{tan2018convolutional}, is an essentially causal CED architecture with two LSTM layers between the encoder and the decoder. Here, LSTM is specifically used to model the temporal dependencies. The encoder consists of five Conv2d blocks aiming at extracting high-level features from the input features, or reducing the resolution. Subsequently, the decoder reconstructs the low-resolution features to the original size of the input, leading the encoder-decoder structure to a symmetric design. In detail, the encoder/decoder Conv2d block is composed of a convolution/deconvolution layer followed by batch normalization and activation function. Skip-connection is conducive to flowing the gradient by concentrating the encoder and decoder.

Unlike the original CRN with magnitude mapping, Tan \textit{et al.}~\cite{tan2019complex} recently proposed a modified structure with one encoder and two decoders to model the real and imaginary parts of complex STFT spectrogram from the input mixture to clean speech. Compared with the traditional magnitude-only target, enhancing magnitude and phase simultaneously has obtained remarkable improvement. However, they treat real and imaginary parts as two input channels, only applying a real-valued convolution operation with one shared real-valued convolution filter, which is not confined with the complex multiply rules. Hence the networks may learn the real and imaginary parts without prior knowledge. To address this issue, in this paper, the proposed DCCRN modifies CRN substantially with complex CNN and complex batch normalization layer in encoder/decoder, and complex LSTM is also considered to replace the traditional LSTM. Specifically, the complex module models the correlation between magnitude and phase with the simulation of complex multiplication.

\vspace{-0.15cm} 
\begin{figure}[ht]
  \centering
  \includegraphics[width=8cm]{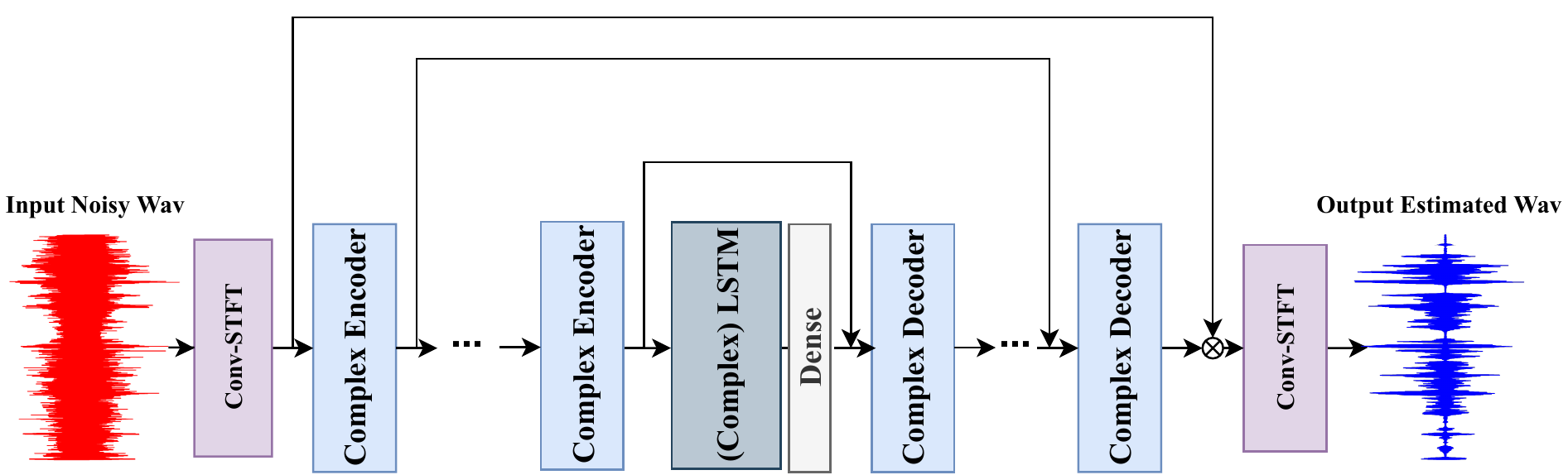} 
  \vspace{0.1cm}
  \caption{DCCRN network}
  \label{fig:dccrn}
  \vspace{-0.5cm} 
\end{figure}

\subsection{Encoder and decoder with complex network}
\begin{figure}[h]
\subfigure[complex convolution]{
    \begin{minipage}[t]{0.5\linewidth}
        \centering
        \includegraphics[width=3cm]{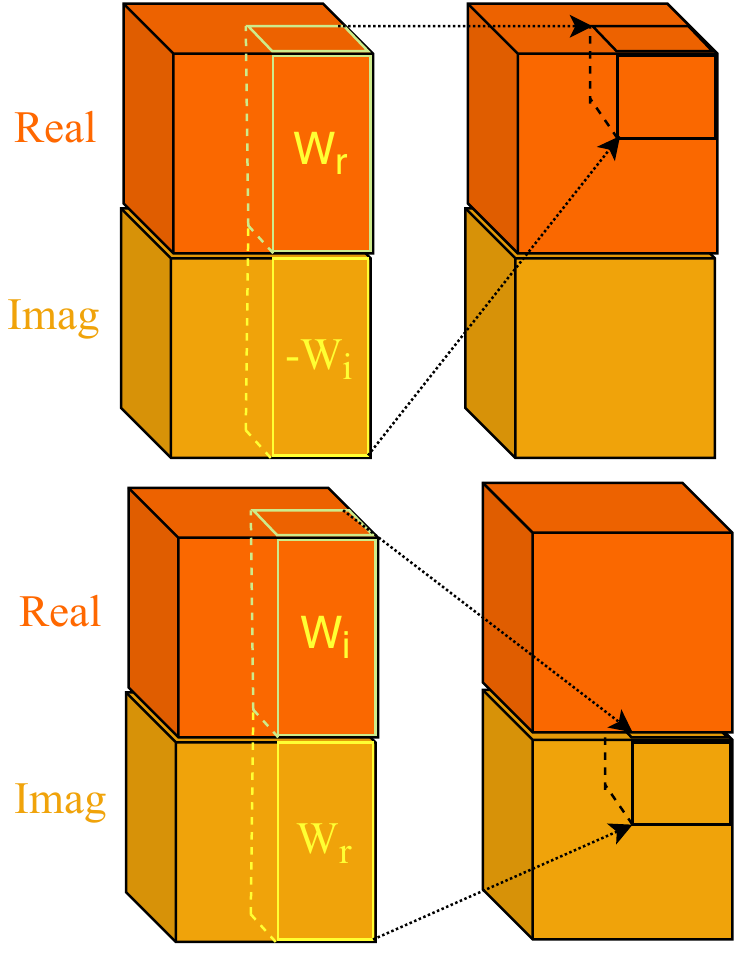}\\
        \vspace{0.02cm}
    \end{minipage}%
}%
\subfigure[complex encoder]{
    \begin{minipage}[t]{0.5\linewidth}
        \centering
        \includegraphics[width=3cm]{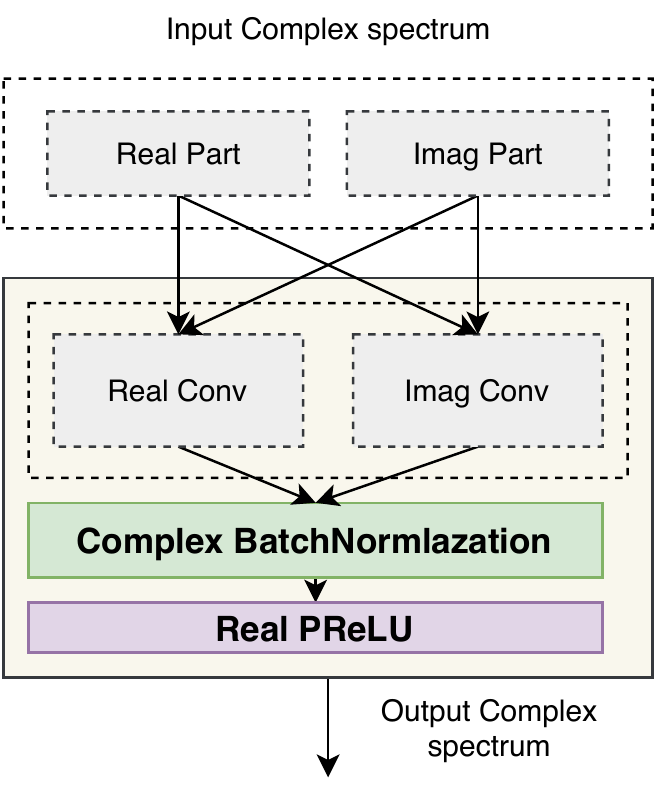}\\
        \vspace{0.02cm}
    \end{minipage}%
}%
\caption{Complex module}
\end{figure}
The complex encoder block includes complex Conv2d, complex batch normalization~\cite{trabelsi2017deep} and real-valued PReLU~\cite{he2015delving}. The complex batch normalization and PReLU follow the implementation of the original paper. We design the complex Conv2d block according to that in DCUNET\cite{choi2019phase}. Complex Conv2d consists of four traditional Conv2d operations, which control the complex information flow throughout the encoder. The complex-valued convolutional filter $W$ is defined  as
$W = W_{r}+j W_{i}$, where the real-valued matrices $W_{r}$ and $W_{i}$ represent the real and imaginary part of a complex convolution kernel, respectively. 
At the same time, we define the input complex matrix $X = X_{r}+j X_{i}$ . Therefore, we can get complex output $Y$ from the complex convolution operation $X\circledast  W$:
\begin{equation}
    F_{out}  = (X_{r} * W_{r}-X_{i} * W_{i})+j(X_{r}* W_{i}+X_{i}* W_{r})
    \label{eq1}
\end{equation}
where $F_{out}$ denotes the output feature of one complex layer.


Similar to complex convolution, given the real and imaginary parts of the complex input $X_r$ and $X_i$, complex LSTM output $F_{out}$ can be defined as:
\begin{eqnarray}
	&&F_{rr} =  \text{LSTM}_r(X_r);\quad F_{ir} = \text{LSTM}_r(X_i) \\
    &&F_{ri} =  \text{LSTM}_i(X_r);\quad F_{ii} = \text{LSTM}_i(X_i) \\
    &&F_{\text{out}}  =  (F_{rr} - F_{ii})+j(F_{ri}+F_{ir})
\end{eqnarray}
where $\text{LSTM}_r$ and $\text{LSTM}_i$ represent two traditional LSTMs of real part and imaginary part, and $F_{ri}$ is caculated by
input $X_r$ with $\text{LSTM}_i$.

\subsection{Training target}



When training, DCCRN estimates CRM and is optimized by signal approximation (SA). Given the complex-valued STFT spectrogram of clean speech $S$ and noisy speech $Y$, CRM can be defined as 
\begin{equation}
    \text{CRM} = \frac{Y_r S_r+Y_i S_i}{Y_r^2+Y_i^2} + j \frac{Y_r S_i-Y_i  S_r}{Y_r^2+Y_i^2}
\end{equation}
where $Y_r$ and $Y_i$ denote the real and imaginary parts of the noisy complex spectrogram, respectively. The real and imaginary parts of the clean complex spectrogram are represented by $S_r$ and $S_i$. 
Magnitude target SMM also can be used for comparison: $\text{SMM} = \frac{|S|}{|Y|}$, where $|S|$ and $|Y|$ indicate the magnitude of clean speech and noisy speech, respectively.
We apply signal approximation, which directly minimizes the difference between the magnitude or complex  spectrogram of clean speech and that of noisy speech applied with mask. The loss function of SA becomes $\text{CSA} = Loss({\Tilde{M}}\cdot{Y}, S)$ and $\text{MSA} = Loss(\Tilde{|M|}\cdot|Y|, |S|) $, where CSA and MSA denote the CRM-based SA and SMM based SA, respectively. Alternatively, the Cartesian coordinate representation $\Tilde{M}=\Tilde{M_r}+j\Tilde{M_i}$ can also be expressed in polar coordinates:
\renewcommand{\arraystretch}{0.1}
\begin{equation}
\left \{
    \setlength{\arraycolsep}{0.2pt}
    \begin{array}{lll}
    \Tilde{M}_{\text{mag}}  &= \sqrt{\Tilde{M_r}^{2}+\Tilde{M_i}^{2}}, &\\
    \Tilde{M}_{\text{phase}} &= \arctan2(\Tilde{M_i},\Tilde{M_r}) &
    \end{array}
\right.
\end{equation}
\renewcommand{\arraystretch}{1.}

We can use three multiplicative patterns for DCCRN, which will be compared with experiments shortly. Specifically, the estimated clean speech $\Tilde{S}$ can be calculated as below.


\begin{itemize}
\vspace{-0.2cm}
\item  DCCRN-R:
\begin{equation}
    \Tilde{S} = (Y_r\cdot\Tilde{M}_r) + j(Y_i\cdot\Tilde{M}_i)
\end{equation}
\item DCCRN-C:
\begin{equation}
    \Tilde{S} = (Y_r\cdot \Tilde{M}_r-Y_i\cdot \Tilde{M}_i)+j(Y_r\cdot \Tilde{M}_i+Y_i\cdot \Tilde{M}_r)
\end{equation}
\item DCCRN-E:
\begin{equation}
    \Tilde{S} = Y_{\text{mag}}\cdot\Tilde{M}_{\text{mag}} \cdot e^{{Y}_{\text{phase}}+\Tilde{M}_{\text{phase}}}
\end{equation}
\vspace{-0.5cm}
\end{itemize}

DCCRN-C obtains $\Tilde{S}$ in the manner of CSA and DCCRN-R estimates the mask of the real and imaginary parts of $\Tilde{Y}$, respectively.
Moreover, DCCRN-E performs in polar coordinates, and it is mathematically similar to DCCRN-C. The difference is that DCCRN-E uses the $tanh$  activation function to limit the mask magnitude to 0 to 1.

\subsection{Loss function}
The loss function of model training is SI-SNR, which has been commonly used as an evaluation metric to replace the mean square error (MSE). SI-SNR is defined as:
\begin{equation}
     \setlength{\arraycolsep}{0.3pt}
     \begin{cases}
    \bm{s}_{\text{target}}&:=(<\Tilde{\bm{s}},\bm{s}>\cdot \bm{s})/||\bm{s}||_{2}^{2} \\
     \bm{e}_{\text{noise}}&:=\bm{\Tilde{s}}-\bm{s_{\text{target}}}\\
     \text{SI-SNR} &:=10\log10(\dfrac{||\bm{s_{\text{target}}}||_{2}^{2}}{||\bm{e_{\text{noise}}}||_{2}^{2}}) \\
     \end{cases}
\end{equation}
where $\bm{s}$ and $\Tilde{\bm{s}}$ are the clean and estimated time-domain waveform, respectively. $<\cdot,\cdot>$ denotes the dot product between two vectors and $||\cdot||_2$ is Euclidean norm (L2 norm). In details, we use STFT kernel initialized convolution/deconvolution module to analyze/synthesize waveform \cite{gu2019end} before sending to network and calculating the loss function.

\section{Experiments}

\subsection{Datasets}


In our experiments, we first evaluated the proposed models as well as several baselines on a dataset simulated on WSJ0~\cite{garofolo1993csr}, and then the best-performed models were further evaluated on the Interspeech2020 DNS Challenge dataset~\cite{reddy2020interspeech}. For the first dataset, we select 24500 utterances (about 50 hours) from WSJ0~\cite{garofolo1993csr}, which includes 131 speakers (66 males and 65 females). We shuffle and split training, validation, and evaluation sets to 20000, 3000 and 1500 utterances, respectively. The noise dataset contains 6.2 hours free-sound noise and 42.6 hours music from MUSAN~\cite{musan2015}, which we use 41.8 hours for training and validation, and the rest 7 hours for evaluation. The speech-noise mixtures in training and validation are generated by randomly selecting utterances from the speech set and the noise set and mixing them at random SNR between -5 dB and 20 dB. The evaluation set is generated at 5 typical SNRs (0 dB, 5 dB, 10 dB, 15 dB, 20 dB).

The second big dataset is based on the data provided by the DNS challenge. The 180-hour DNS challenge noise set includes 150 classes and 65,000 noise clips and the clean speech set includes over 500 hours of clips from 2150 speakers. To make full use of the dataset, we simulate the speech-noise mixture with dynamic mixing during model training. In detail, at each training epoch, we ﬁrst convolve speech and noise with a room impulse response (RIR) randomly-selected from a simulated 3000-RIR set by the image method \cite{allen1979image}, and then the speech-noise mixtures are generated dynamically by mixing 
reverb speech and noise at random SNR between -5 and 20 dB. The total data `seen' by the model is over 5000 hours after 10 epochs of training. We use the official test set for objective scoring and final model selection.

\subsection{Training setup and baselines}
For all of the models, the window length and hop size are 25 ms and 6.25 ms, and the FFT length is 512. We use Pytorch to train the models, and the optimizer is Adam. The initial learning rate is set to 0.001, and it will decay 0.5 when the validation loss goes up. All the waveforms are resampled at 16k Hz. The models are selected by early stopping.
In order to choose the model for the DNS challenge, we compare several models on the WSJ0 simulation dataset, described as follows.
\vspace{-0.cm}
\begin{description}
\item{LSTM}: a semi-causal model contains two LSTM layers, and each layer has 800 units; we add one Conv1d layer in which kernel size is 7 in the time dimension, and the look-ahead is 6 frames to achieve semi-causal. The output layer is a 257-unit fully-connected layer. The input and output are the noisy and estimated clean spectrogram with MSA, respectively.

\item{CRN}: a semi-causal model contains one encoder and two decoders with the best configuration in \cite{tan2019complex}. The input and output are the real and imaginary part of the noisy and estimated STFT complex spectrogram. Two decoders process the real and imaginary parts separately. The kernel size is also (3,2) in frequency and time dimension, and the stride is set to (2,1). For the encoder, we concatenate real and imaginary parts in the channel dimension, so the shape of 
the input feature is [BatchSize, 2, Frequency, Time]. Moreover, the output channel of each layer in encoder is \{16,32,64,128,256,256\}. The hidden LSTM units are 256, and a dense layer with 1280 units is after the last LSTM. On account of skip connection, each layer in input channel of real or imaginary decoder  is \{512,512,256,128,64,32\}.

\item{DCCRN}: four models consist of DCCRN-R, DCCRN-C, DCCRN-E and DCCRN-CL (masking like DCCRN-E).
The direct current component of all these models is removed. The number of channel for the first three DCCRN is \{32,64,128,128,256,256\}, while the DCCRN-CL is \{32,64,128,256,256,256\}. The kernel size and stride are set to (5,2) and (2,1), respectively. The real LSTMs of the first three DCCRN are two layers with 256 units and DCCRN-CL uses complex LSTM with 128 units for the real part and imaginary part, respectively. And a dense layer with 1024 units is after the last LSTM. 

\item{DCUNET}: we use DCUNET-16 for comparison and the stride in time dimension is set to 1 to fit with the DNS challenge rules. Moreover, the channels in encoder is set to [72,72,144,144,144,160,160,180].

\end{description}For the implementation of semi-causal convolution\cite{bahmaninezhad2019unified}, there are only two differences with commonly used causal convolution in practice. First, we pad zeros in front of the time dimension at each Conv2ds in the encoder. Second, for the decoder, we look ahead one frame in each convolution layer. This eventually leads to 6 frames look-head, totally $6\times6.25=37.5$ ms, confined with the DNS challenge limit --- 40 ms.

\subsection{Experimental results and discussion}
The model performance is first assessed by PESQ\footnote[1]{\url{https://www.itu.int/rec/T-REC-P.862-200102-I/en}} on the simulated WSJ0 dataset. Table~\ref{tab:semiwsj} presents the PESQ score on the test sets. 
In each case, the best result is highlighted by a boldface number.
\setlength{\abovecaptionskip}{0pt} 
\setlength{\belowcaptionskip}{0pt}
\setlength{\abovedisplayskip}{0pt}
\setlength{\belowdisplayskip}{0pt}
\vspace{-0.2cm}
\begin{table}[!h]
\setlength\tabcolsep{2.5pt}
\footnotesize
\caption{PESQ on the simulated WSJ0 dataset}
\begin{tabular}{lclllllll}
\toprule 
Model& Para.(M)&0dB&5dB&10dB& 15dB&20dB&Ave.\\  
\midrule
Noisy&  -    &2.062&2.388&2.719&3.049&3.370&2.518\\  
LSTM&9.6&2.783&3.103&3.371&3.593&3.781&3.326\\  
CRN &6.1&2.850&3.143&3.374&3.561&3.717&3.329\\ \midrule 
DCCRN-R&3.7&2.832&3.192&3.488&3.717&3.891&3.424\\  
DCCRN-C&3.7&2.832&3.187&3.477&3.707&3.840&3.409\\  
DCCRN-E&3.7&2.859&3.203&3.492&3.718&3.891&3.433\\  
DCCRN-CL&3.7&\textbf{2.972}&\textbf{3.301}&\textbf{3.559}&3.755&3.901&3.498\\ \midrule
DCUNET &3.6&2.971&3.297&3.556&\textbf{3.760}&\textbf{3.916}&\textbf{3.500} \\ 
\midrule
\end{tabular}
\label{tab:semiwsj}
\vspace{-0.3cm}
\end{table}

On the simulated WSJ0 test set, we can see that the four DCCRNs outperform the baseline LSTM and CRN, which indicates the effectiveness of complex convolution. DCCRN-CL achieves better performance than other DCCRNs. This further shows that complex LSTM is also beneficial to complex target training. Moreover, we can see that full-complex-value network DCCRN and DCUNET are similar in PESQ. It worth noting that the computational complexity of DCUNET is almost 6 times than that of DCCRN-CL, according to our run-time test.


\vspace{0.1cm}
\begin{table}[htbp]
\footnotesize
\setlength\tabcolsep{2pt}
\caption{PESQ on DNS challenge test set (simulated data only). T1 and T2 denote track 1 (real-time-track) and track 2 (non-real-time-track).}
\vspace{0.2cm}
\begin{tabular}{lccccc}
\toprule 
Model                         & \tabincell{c}{Para.\\(M)} & \tabincell{c}{look-ahead\\(ms)}          & no reverb  & reverb  & Ave.\\ 
\midrule 
Noisy           & -         & -               & 2.454     & 2.752   & 2.603\\ 
NSNet (Baseline)~\cite{9054254} &  1.3     & 0            & 2.683     & 2.453   & 2.568\\ 
DCCRN-E {[}T1{]}              &  3.7      & 37.5         & \textbf{3.266}     & 3.077   & 3.171\\ 
DCCRN-E-Aug {[}T2{]} &  3.7     & 37.5         & 3.209     &  \textbf{3.219}  & \textbf{3.214} \\ 
DCCRN-CL {[}T2{]}             &  3.7     & 37.5         & 3.262     &  3.101  & 3.181       \\
DCUNET {[} T2{]}                 &  3.6     & 37.5         & 3.223     &  2.796  & 3.001       \\ 
\bottomrule
\label{tab:dnspesq}
\end{tabular}
\vspace{0.1cm}
\setlength\tabcolsep{2pt}
\caption{MOS on DNS challenge blind test set \text{\cite{reddy2020interspeech}}}
\vspace{0.2cm}
\begin{tabular}{llcccccc}
\toprule
\multicolumn{2}{l}{Model}             & Para.(M)           & no reverb     & reverb        & realrec        & Ave.  \\ 
\midrule 
\multicolumn{2}{l}{Noisy}            &       -          &  3.13              &     2.64      &    2.83      &  2.85     \\ 
\multicolumn{2}{l}{NSNet (Baseline)~\cite{9054254} }         &      1.3         &  3.49              &     2.64      & 3.00              &   3.03    \\  
\midrule
\multirow{3}*{\shortstack{Track 1}} & DCCRN-E    &      3.7         &  \textbf{4.00}     &     2.94      &  \textbf{3.37}    & \textbf{3.42}      \\ 
& Team 9              &     UNK          &  3.87              & \textbf{2.97} &     3.28          &  3.39        \\ 
&Team 17            &     UNK          &  3.83              & 3.05          &      3.27         & 3.34 \\     
\midrule
\multirow{3}*{\shortstack{Track 2}}&Team 9         &     UNK          & \textbf{4.07}      &\textbf{3.19}  & \textbf{3.40} & \textbf{3.52} \\
&DCCRN-E-Aug  &      3.7         &  3.90              &     2.96      &   3.34               & 3.38 \\ 
& Team 17            &     UNK          &   3.83             &     3.15      &   3.28                &  3.38 \\
\midrule
\end{tabular}
\label{tab:dnsmos}
\vspace{-0.5cm}
\end{table}

In the DNS challenge, we evaluate the two best DCCRN models and DCUNET with the DNS dataset. Table~\ref{tab:dnspesq} shows the PESQ scores on the test set.
Similarly, DCCRN-CL achieves a little bit better PESQ than DCCRN-E in general. But after our internal subject listening, we find DCCRN-CL may over-suppress the speech signal on some clips, leading to unpleasant listening experiences. DCUNET obtains relatively good PESQ on the synthetic non-reverb set, but its PESQ will drop significantly on the synthetic reverb set. We believe that subjective listening becomes very critical when the objective scores are close for different systems. For these reasons, DCCRN-E was finally chosen for the real-time track. In order to improve the performance on the reverb set, we add more RIRs in the training set to result in a model called DCCRN-E-Aug, which was chosen for the non-real-time track. According to the results on the final blind test set in Table~\ref{tab:dnsmos}, the MOS of DCCRN-E-Aug has a small improvement of 0.02 on the reverb set. Table~\ref{tab:dnsmos} summarizes the final P.808 subjective evaluation results for several top systems in both tracks provided by the challenge organizer. We can see that our submitted models perform well in general. DCCRN-E achieves an average MOS of 3.42 on all sets and 4.00 on the non-reverb set. 
The one frame processing time of our PyTorch implementation of DCCRN-E (exported by ONNX) is 3.12 ms tested empirically on an Intel i5-8250U PC.
Some of the enhanced audio clips can be found from \url{https://huyanxin.github.io/DeepComplexCRN}.

\section{Conclusions}
In this study, we have proposed a deep complex convolution recurrent network for speech enhancement. The DCCRN model utilizes a complex network for complex-valued spectrum modeling. With the complex multiply rule constraint, DCCRN can achieve better performance than others in terms of PESQ and MOS in the similar configuration of model parameters. In the future, we will try to deploy DCCRN in low computational scenarios like edge devices. We will also enable DCCRN with improved noise suppression ability in reverberation conditions.

\bibliographystyle{IEEEtran}

\bibliography{reference}

\end{document}